\documentclass[
reprint,
superscriptaddress,
groupedaddress,
preprintnumbers,
bibnotes,
 amsmath,amssymb,
 aps,
]{revtex4-2}
\usepackage{float}
\usepackage{graphicx}
\usepackage{dcolumn}
\usepackage{bm}

\begin{document}

\title{Enhancement of dynamical coupling in artificial spin-ice systems by incorporating perpendicularly magnetized ferromagnetic matrix
}

\author{Syamlal Sankaran Kunnath}
\email{syamlal.sankaran@amu.edu.pl}
\author{Mateusz Zelent}

\author{Mathieu Moalic}

\author{Maciej Krawczyk}
  
\affiliation {Institute of Spintronics and Quantum Information, Faculty of Physics and Astronomy, Adam Mickiewicz University, Poznan, 61-614 Poznan, Poland}

\date{\today}

\begin{abstract}
Artificial spin-ice systems, consisting of arrays of interacting ferromagnetic nanoelements, offer a versatile platform for reconfigurable magnonics with potential in GHz logic and neuromorphic computing. However, weak dipolar coupling between nanoelements severely limits their functionality.
We numerically demonstrate a rich spin-wave spectrum in a square spin-ice structure immersed in a perpendicularly magnetized ferromagnetic matrix, which is different from a single spin-ice system. We observe a strong magnon-magnon coupling between the bulk second-order mode of the nanoelements and the fundamental mode of the matrix, supported by a pronounced anticrossing frequency gap. We show that, in addition to the dipolar coupling, exchange interactions at the nanoelement-matrix interface play a crucial role in this hybridization. Furthermore, the strength of the coupling can be enhanced by almost 40\% just by reconfiguring the magnetization at the vertices from low-energy to high-energy monopole states. These results open the way to exploit artificial spin-ice systems for magnonic applications, taking advantage of the strong coupling and vertex-dependent dynamics.
\end{abstract}


\maketitle

\section{Introduction}

Spin waves (SWs) are being explored for information transfer and processing in reconfigurable
systems~\cite{krawczyk2014review,kruglyak2010magnonics}. They offer numerous advantages as data carriers, including reduced waste heat, tunable wavelengths down to sub-100 nm, frequencies ranging from below 1 GHz to tens of THz, and coherent coupling between different SW modes (magnons), photons, and phonons~\cite{barman2018spin,barman20212021,chumak2022advances,flebus20242024}. Recently, reconfigurable magnonic crystals (RMCs), a class of metamaterials, in which the magnetization state can be changed, have made significant progress in achieving useful functionalities, in particular, for low-power computing, high-speed data processing, and advanced signal processing~\cite{haldar2016reconfigurable,krawczyk2014review,barman2020magnetization,grundler2015reconfigurable,chumak2014magnon,Ji2022,Szulc2022}. 

Artificial spin ice (ASI) systems, which can be considered as a subset of RMCs, feature inter-element couplings, primarily governed by long-range magnetic dipole interactions, which introduce geometrical frustration of the magnetization orientation in a mono-domain ferromagnetic nanoelments~\cite{heyderman2013artificial,nisoli2013colloquium,lendinez2019magnetization, gilbert2014emergent}.
As a result, the ASI system exhibits a rich variety of magnetic states and behaviors, including the formation of magnetic monopole-like excitations, extensive degeneracy, Coulomb phase and other complex phenomena~\cite{ladak2010direct,king2021qubit,branford2012emerging,drisko2017topological,may2021magnetic}. 
The resulting emergent magnetic monopoles and ground state magnetization degeneracy, combined with exceptional flexibility in lattice design, make ASI a promising candidate for reprogrammable functional magnonic devices~\cite{gliga2020dynamics,nisoli2017deliberate, skjaervo2020advances}. Thus, magnetic resonances and SW confined modes 
have been investigated in a variety of ASI systems in recent times~\cite{barman2020magnetization,grundler2015reconfigurable,petti2022review}. Collective SW behaviour, mode hybridization, and nonlinear multimagnon scattering have been demonstrated, highlighting the intricate and dynamic nature of these systems
~\cite{gliga2013spectral,jungfleisch2016dynamic,dion2022observation,lendinez2021emergent,lendinez2023nonlinear,lendinez2021observation,gliga2015broken}. The evolution of mode dynamics from the ground state to charged monopole states has also been of particular interest, as investigated in depth in Ref.~\cite{gliga2013spectral}.

Moreover, recent research suggests that exploring the range of accessible microstates and unique magnonic behaviors in ASI systems can significantly enhance the capabilities for reservoir computing, a subset of neuromorphic computing~\cite{manneschi2024optimising}. 
In addition to the accessible magnetic states, the ASIs are also explored for magnon-magnon coupling~\cite{macneill2019gigahertz,gartside2018realization,pal2024using,dion2024ultrastrong, mondal2024brillouin}. 
 Specifically, mode hybridizations and frequency anticrossing are highly dependent on vertex types in square ice configurations, with charged monopole vertices exhibiting an enhanced frequency gap~\cite{gartside2021reconfigurable}. Some studies have tuned magnon coupling strength in ASI through controlled dipolar interactions, interlayer exchange interaction, and modifying external magnetic field direction~\cite{dion2024ultrastrong}.
However, to date, ASI systems, even when based on multilayer ferromagnetic nano-elements, rely on a relatively weak, long-range dipolar coupling between them, which severely limits the strength of the coupling and the feasibility of the system to transfer information between discrete islands. These limit the potential applications of ASIs in magnonics.

Our previous experimental and numerical studies have demonstrated the SW dynamics in two-dimensional magnonic antidot lattice (ADLs) based on perpendicular magnetic anisotropy (PMA) multilayers~\cite{pan2020edge,moalic2024role,pal2012time}. In these systems, when the PMA multilayer is patterned with Ga ions (in a FIB process), the PMA at the edges of the antidots is significantly reduced. This results in the formation of in-plane magnetized edges around the antidots, which are immersed in an out-of-plane magnetized matrix, in remanence, or small biased magnetic fields. The recent results indicate a rather strong magnon-magnon coupling between the azimuthal SW mode in the rim and the SW mode of the PMA matrix. Interestingly, the exchange interactions at the rim-PMA matrix interface have been identified as a key source of this coupling. 
The coexistence of two magnetically different, in-plane and out-of-plane magnetized, but strongly coupled subsystems in the multilayer opens the possibility of creating a novel type of ASI system.
Therefore, we propose to use square lattices of nanoelements with reduced PMA immersed in the metallic multilayer with interface-induced PMA as an ASI.

We investigate the SW dynamics in in-plane magnetized ferromagnetic nanoelements forming a square lattice ASI (SSI), and integrated with perpendicularly magnetized Co/Pd multilayer (SSI-PMA) using micromagnetic simulations. We observe rich and distinct magnonic spectra in SSI-PMA structures compared to the standalone SSI system. In particular, the second-order bulk mode of the SSI nanoelement hybridizes with the fundamental bulk mode of the PMA matrix, resulting in significant anticrossing frequency gaps and thus strong magnon-magnon coupling. By varying the vertex gap between the nanoelements and adjusting the effective thickness of the multilayer, the coupling strength can be tuned. Importantly, our results reveal that mode hybridization and coupling strength depend on the vertex type of the SSI, allowing the variety of magnetization states in ASIs to be exploited to control magnon-magnon coupling and collective SW dynamics.
The proposed system can be fabricated experimentally by controlled focused ion irradiation of ferromagnetic multilayers with PMA and the predicted properties can be measured with a standard broadband ferromagnetic resonance technique. We have therefore shown that the study of PMA-based ASI systems is an important and compelling direction for future research in magnetism and magnonics.

\section{Structure and Methods \label{Sec:Methods}}

\begin{figure*}
    \centering
    \includegraphics[width=0.95\linewidth]{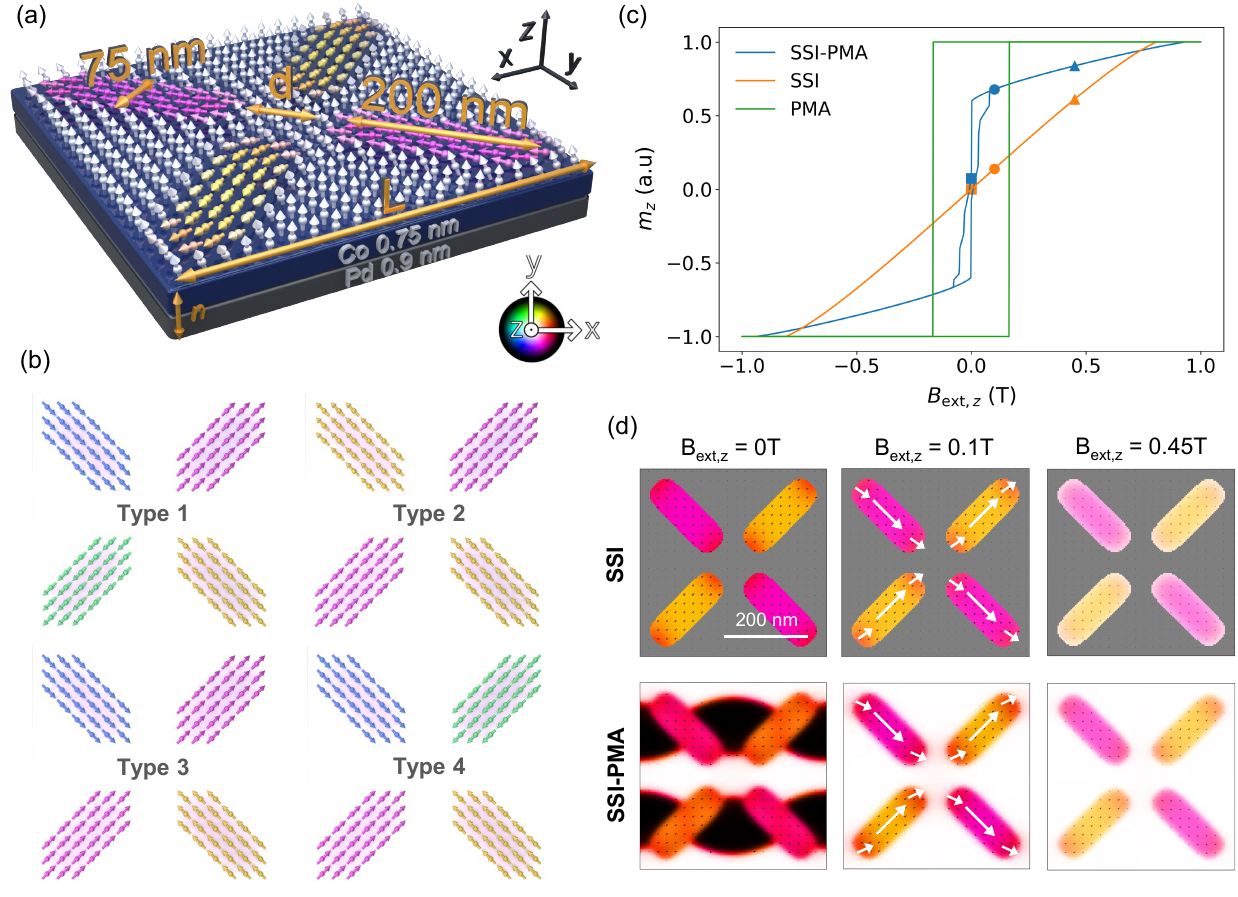}
    \caption{\textbf{Schematic illustration of the structure and its static magnetization configurations}: (a) Schematic of the unit cell of the SSI-PMA architecture, showing in-plane magnetized nanoelements that are 200 nm long and 75 nm wide, configured as a square ice lattice and embedded in a perpendicularly magnetized matrix. In the study, the lattice constant $L$ is varied according to the vertex gap between the nanoelements $d$. The number of Co/Pd multilayer repetitions, $n$, and so the multilayer thickness are also varied. (b) Schematic representation of the four possible types of magnetization arrangement in the SSI vertex. Type 1 and Type 2 follow the ice rule and are low energy states, while Type 3 and Type 4 host two ($+2Q$) and four ($+4Q$) magnetic monopoles, respectively. (c) Combined plot showing the static magnetization response versus the out-of-plane bias magnetic field for SSI, PMA, and SSI-PMA structures with the same lattice constant. (d) The corresponding magnetization configurations at remanence, 0.1 T and 0.45 T for the SSI and SSI-PMA systems.}
    \label{fig1}
\end{figure*}

Our artificial SSI-PMA system is composed of in-plane magnetized nanoelements of elongated shape with rounded corners, with dimensions of 200 nm length and 75 nm width, immersed in a perpendicularly magnetized matrix, as schematically shown in Fig. \ref{fig1}(a). These dimensions are chosen to closely correspond to those in previous SSI studies~\cite{lendinez2023nonlinear,gliga2013spectral, lendinez2021emergent}. Initially, the lattice constant of the structure $L$ is set to 424.26 nm (diagonal length 600~nm), and a vertex gap between nanoelements $d$ is 100~nm.
The whole film is composed of a multilayer consisting of Co (0.75 nm) and Pd (0.9 nm), with total thickness of 13.2 nm. However, in micromagnetic simulation, this stack is represented as a single ferromagnetic layer with effective properties~\cite{moalic2024role,pan2020edge}. 
The effective material parameters chosen for the Co/Pd multilayer are as follows: exchange constant $A_{\text{ex}} = 13$~pJ/m, and saturation magnetization $M_{\text{s}} = 810$~kA/m. For clarity of the spectra in most simulations, we assume a negligible Gilbert damping constant $\alpha$, for selected cases a realistic value $\alpha=0.008$ is assumed. We assume a bulk magnetic anisotropy, with constant $K_\text{u,bulk} = 4.5 \times 10^{5}$~J/m³ along the $z$ axis in the matrix region, while it is set to 0 in the in-plane magnetized SSI nanoelements.  

The micromagnetic simulations are performed using own version of $Mumax^{3}$~\cite{vansteenkiste2014design}, called $Amumax$~\cite{amumax2023}. More detailed description of simulation procedures can be found in the
Supplemental Information Sec. I.

 \section{\label{sec:level3}Results and Discussion}
\subsection{Artificial square spin ice and PMA}

\begin{figure*}
    \centering
    \includegraphics[width=0.95\linewidth]{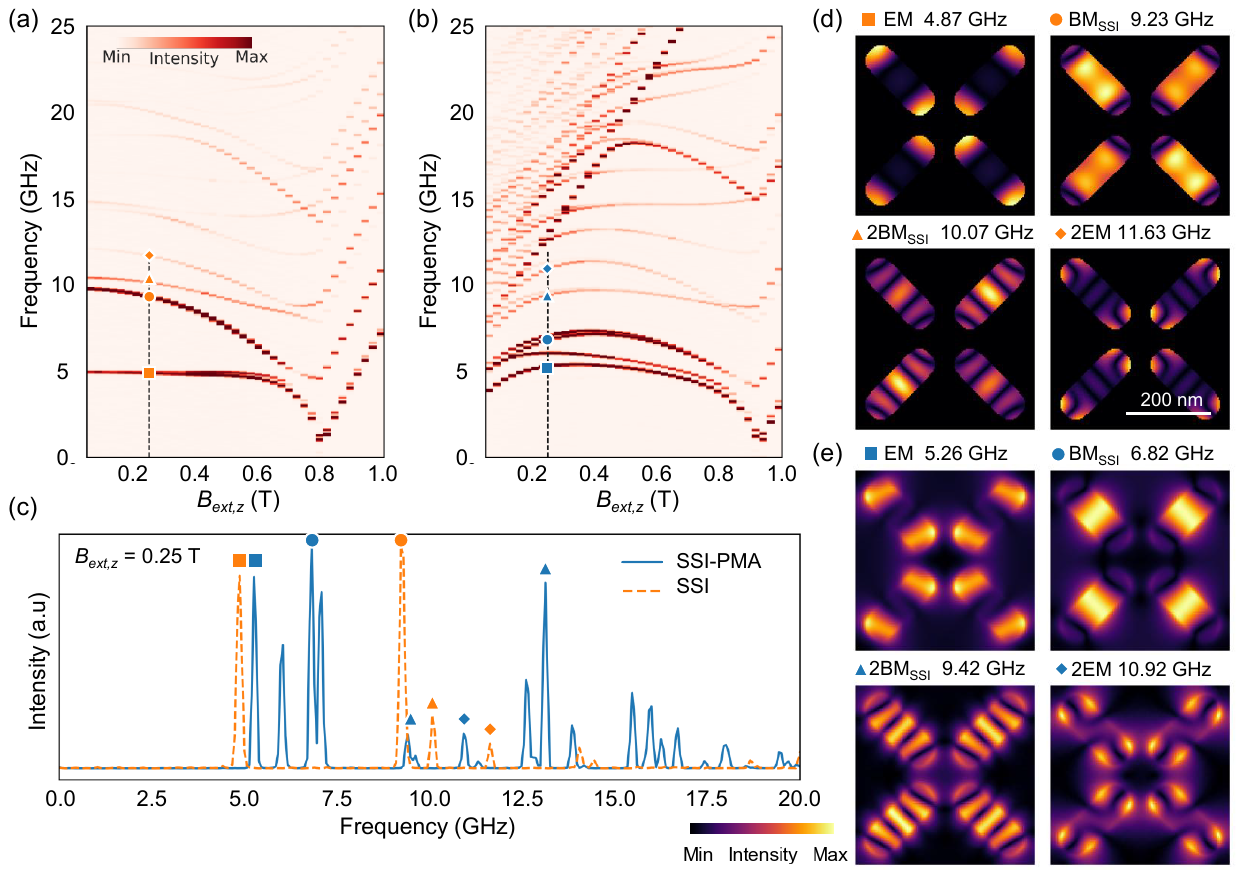}
    \caption{\textbf{SW resonance spectra and mode dynamics in SSI and SSI-PMA systems in Type 2 magnetization arrangement}: (a,b) Evolution of the SW resonance spectra for SSI (a) and SSI-PMA (b) systems as a function of the applied bias magnetic field along the out-of-plane direction. The vertical black dashed line indicates the field value 0.25 T for which profiles of SW amplitude are plotted in (d)-(e). (c) FFT spectra of SSI and SSI-PMA at field value of 0.25 T. Snapshots of EM, $\textbf{BM}_{\mathrm{SSI}}$ and their second order modes at 0.25 T for the SSI (d) and SSI-PMA (e) system. The modes are clearly marked on the (a)-(c) spectra.}
    \label{fig2}
\end{figure*}

Our SSI-PMA consists of periodic vortices, connected by a PMA material, where four nanoelement magnetic moments converge. These vertices are generally categorized into four different types based on their magnetization configurations, as illustrated in Fig.~\ref{fig1}(b). Each nanoelement in the vertex has two magnetization states that either point in or out. Out of the 16 possible configurations, 6 satisfy the two-in-two-out ice rule, corresponding to Type 1 and Type 2 vertices. The remaining configurations are classified as Type 3 or Type 4 vertices and host magnetic monopole charges, $Q=2$ and $Q=4$, respectively. This is similar to the previously extensively studied square-lattice ASIs~\cite{gliga2020dynamics,heyderman2013artificial}. 
To illustrate the novelty of the SSI-PMA system, we will compare its SW spectra with the system of the same in-plane nanoelements but without PMA matrix, which we will call the SSI system.   

Figure \ref{fig1}(c) shows the simulated hysteresis loop for the SSI-PMA, SSI, and additionally for the homogeneous Co/Pd film with PMA (named as PMA, and having the same thickness as matrix in the SSI-PMA), with the magnetic field applied along the out-of-plane direction, and starting the simulations in full saturation with $B_{\text{ext},z}=1.0$~T. The PMA and SSI-PMA systems show behavior characteristics of easy-axis direction, while the SSI system exhibits a hard-axis type hysteresis loop. The large and different fields required for saturation of domains in SSI (0.80~T) and SSI-PMA (0.93~T) are due to the strong dipolar coupling between the in-plane nanoelements, and between nanoelements and PMA matrix, respectively. At remanence and $\pm z$ field values up to the saturation, the in-plane nanoelements in both systems stabilize with Type 2 orientation of magnetization, as illustrated in Fig. \ref{fig1}(d) for $B_\text{ext,z} = 0$, 0.1 and 0.45 T. Although the Type 1 state also satisfies the ice rule, it is difficult to achieve during remagnetization and is more easily achieved by thermal relaxation\cite{gliga2020dynamics}. In contrast, the Type 2 state is more commonly observed during remagnetization in systems with short interelement distances between neighboring nanoelements, which leads to stronger dipolar interactions that favor this configuration. 
Thus in the following section, we will focus on the Type 2 configuration, and in Sec.~\ref{Sec:Monopole} we will extend our research with the other vertice types to study the influence of magnetic monopoles on SW spectra.

The SW characteristics of SSI and SSI-PMA systems were performed under an external out-of-plane magnetic field in the range of 0 to 1~T. The results are shown for both SSI [Fig.~\ref{fig2}(a)] and SSI-PMA [Fig.~\ref{fig2}(b)] systems, each with a vertex gap $d=100$~nm and $L=424.2$~nm (length along diagonal 600 nm). The color represents the SW mode intensity calculated as described in Sec.~\ref{Sec:Methods}. The SW spectra for the SSI and SSI-PMA structures at an arbitrarily chosen field value of 0.25 T, i.e., cross sections of the spectra in Fig.~\ref{fig2}(a)-(b), are shown in Fig.~\ref{fig2}(c). For completeness, the frequency spectrum of the PMA matrix with the holes instead of nanoelements under an external field is presented in Supplementary Information Sec.~II. 
In the SSI system, for all modes, the frequency of the SWs decreases with increasing $B_\text{ext,z}$ until saturation at 0.8~T. Two types of modes can be distinguished: Edge Modes (EMs) and Bulk Modes ($\textbf{BMs}_{\mathrm{SSI}}$). The EMs are localized at the ends of the nanoelements, where demagnetizing field is strongest, and are observed in the lower frequency region~\cite{gliga2013spectral, jungfleisch2016dynamic}. The $\textbf{BMs}_{\mathrm{SSI}}$ have amplitude concentrated at the center of the nanoelements. At higher frequencies there are also 
higher-order EMs and $\textbf{BMs}_{\mathrm{SSI}}$~\cite{jungfleisch2016dynamic,gliga2013spectral,ghosh2019emergent}. For instance, at 0.25 T, the fundamental (i.e., without phase change in the nanoelement) EM and $\textbf{BM}_{\mathrm{SSI}}$ were detected at 4.87 GHz and 9.23 GHz, respectively, with their second-order modes at 11.63 GHz and 10.07 GHz, respectively (see, Fig.~\ref{fig2}(d)). In contrast to the SSI modes, increasing the magnetic field results in a monotonic increase in the frequency of modes observed in PMA matrix spectrum (Fig.~S1).

When SSI is integrated into a PMA matrix, the spectra become significantly different and more complex as compared to SSI [compare Fig.~\ref{fig2}(a) and (b)]. The frequency of the nanoelement modes initially increases with increasing bias field strength and then decreases to out-of-plane saturation at 0.93~T. In addition, the bulk modes from the PMA matrix enter the spectra and show a linear increase in frequency with increasing field strength (similar to the dependence in the PMA film shown in SI, Fig.~S1), with the fundamental bulk mode of the PMA matrix $\textbf{BM}_{\mathrm{PMA}}$ (13.13~GHz at 0.25~T) and higher order modes at higher frequencies.
These changes in the spectra are
due to the entry of the modes from the PMA matrix and the additional interaction between the spins in the in-plane magnetized nanoelements and the out-of-plane matrix region. This interaction includes both dynamic  and static, which will be studied in the following sections. The static magnetization configuration and its evolution with the magnetic field, e.g. as shown in Fig.\ref{fig1}(d), also plays an important role in the dynamics~\cite{jungfleisch2016dynamic,dion2024ultrastrong}. For example, in the SSI-PMA, we observe two EMs and two $\textbf{BMs}_{\mathrm{SSI}}$ with frequencies clearly separated [at 0.25 T, the EMs are at 5.26 GHz and 6.04 GHz, while the $\textbf{BM}_{\mathrm{SSI}}$ are at 6.82 GHz and 7.02 GHz, see also the mode profiles in Fig.~\ref{fig2}(e)].  As the external field increases, the frequency gap between two EMs and two $\textbf{BMs}_{\mathrm{SSI}}$ decreases, causing them to merge.

\subsection{Magnon-magnon hybridization}

\begin{figure*}
    \centering
    \includegraphics[width=1\linewidth]{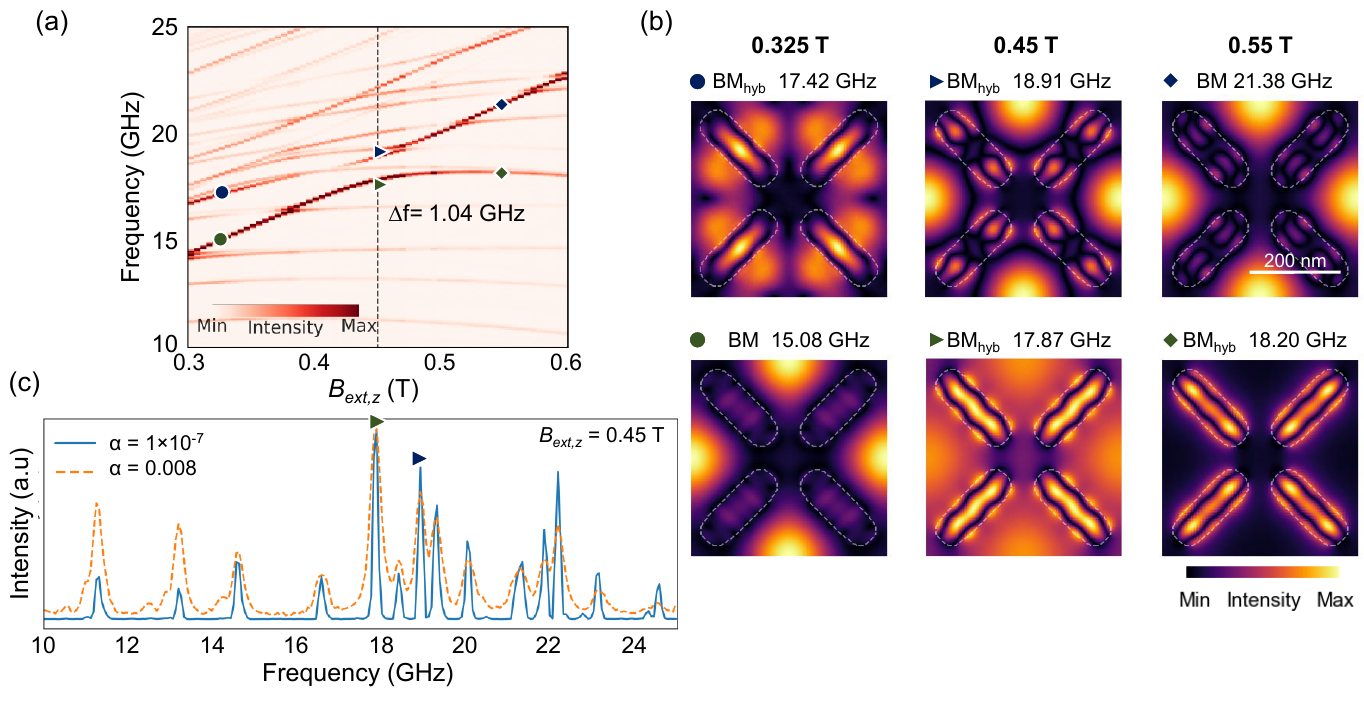}
    \caption{\textbf{Magnon-magnon hybridization in SSI-PMA, Type 2 magnetization arrangement}: (a) A zoom-in part of Fig. \ref{fig2}(b), which shows the evolution of the SW spectra for the SSI-PMA system in dependence on the out-of-plane magnetic field at a higher-frequency range. The vertical dashed line represents the field (0.45~T) at the center of the frequency gap (1.04~GHz). (b) Mode profiles illustrating the evolution and hybridization of $\textbf{BM}_{\mathrm{PMA}}$ with higher-order $\textbf{BM}_{\mathrm{SSI}}$ mode across different magnetic field strengths. For enhanced visibility, the contours of the in-plane nanoelements are highlighted with white dotted lines. (c) FFT spectra for $B_{\text{ext},z} = 0.45$ T with low damping and realistic damping values. 
}
    \label{fig3}
\end{figure*}

Figure~\ref{fig3}(a) shows the zoomed-in view of field-dependent frequency spectra from Fig.~\ref{fig2}(b), revealing key for this study details of the SSI-PMA system's dynamics. Notably, in this frequency range, some higher-order modes from the nanoelements exhibit hybridization with the bulk modes from the PMA matrix ($\textbf{BM}_{\mathrm{hyb}}$). In particular, it happens between the higher-order $\textbf{BM}_{\mathrm{SSI}}$, and the fundamental bulk mode $\textbf{BM}_{\mathrm{PMA}}$ of the matrix. The one centered around 18~GHz at 0.45~T is the strongest among the various hybridizations observed, and it maintains a significant anticrossing frequency gap $\Delta f=1.04$~GHz. The fundamental $\textbf{BM}_{\mathrm{PMA}}$ mode with in-phase magnetization oscillation in the whole matrix, and the 2$^{nd}$ order $\textbf{BM}_{\mathrm{SSI}}$ mode quantized along the short axis of the nanoelement, are involved in this process. The evolution of their profiles with increasing the magnetic field is illustrated in Fig.~\ref{fig3}(b) (see also the phase distribution in SI Fig.~S9). It is clear that during mode hybridization, the collective oscillations involve contributions from the in-plane magnetized SSI elements, the interface between SSI and PMA matrix as well as the PMA bulk region.  Fig. \ref{fig3}(c), whthis considered SS remains evident even with a realistic damping value $\alpha = 0.008$~\cite{pan2020edge}, see the spectra in Fig.~\ref{fig3}(c), emphasizing the relevance of these findings to practical applications. Additional details regarding the frequency versus field spectra and the corresponding modes are elaborated in Supplementary Information, Sec.~VI.

To elucidate the type of interaction responsible for this hybridization, we introduce a small non-magnetic spacer layer (dimension is about 2 discretization cells) between the SSI nanoelements and the PMA matrix. This breaks off the exchange interactions between the nanoelements and the PMA matrix.  As shown in SI Sec. III, this results in a weak mode hybridization with a small anticrossing frequency gap, $\Delta f = 0.32$~GHz. This indicates that the observed magnon-magnon coupling, similar to Ref.~\cite{moalic2024role}, is primarily driven by exchange interactions between the in-plane magnetized nanoelements and the PMA region. This is in contrast to the previous reports~\cite{dion2022observation,shen2022dipolar,dion2024ultrastrong}, where poor dipolar coupling was responsible for the strong magnon-magnon coupling. 

Nevertheless, the dipolar interactions also contribute to the mode coupling in the SSI-PMA system. This is indicated by the 0.32~GHz anti-crossing gap when the exchange coupling between the SSI nanoelements and the PMA matrix is broken, but even more meaningful is the influence of the neighbouring nanoelements on the gap width.  We analyzed the frequency-field diagram and mode profiles for both a single nanoelement and two dipolar-coupled nanoelements surrounded by PMA matrix in the unit cell, with exchange coupled SSI nanoelements and PMA matrix (detailed results of these analyses are presented in SI, Sec. IV). We found that the considered hybridization occurs even for a single nanoelement in the unit cell but with a small gap $\Delta f = 0.45$~GHz. But, when the dipolar coupling between the two nanoelements is present, the gap increases significantly up to $\Delta f = 0.78$~GHz (both at a similar frequency range but a higher magnetic field value 0.5 ~T), and finally to $\Delta f = 1.04$~GHz in a full SSI-PMA system. This suggests that while the exchange interactions alone are sufficient to induce coupling between the modes, the addition of dipolar coupling between the nanoelements significantly enhances it, promising tunability of magnon-magnon coupling and collective SW dynamics over large distances.

To estimate a possibility for experimental observation of our findings, we analyze the average FFT spectra for microwave fields oriented along the $x$- and $z$-axes (see SI, Sec. VIII), corresponding to broadband FMR coplanar waveguide and ring geometry setups, respectively~\cite{dion2024ultrastrong, gartside2021reconfigurable, jungfleisch2016dynamic, ghosh2019emergent}. Simulations reveal hybridized modes in both configurations (see Fig. S8). Notably, for the $x$-axis oriented microwave magnetic field, a significantly stronger absorption is expected due to the more effective excitation of the in-plane magnetized nanoelements in the SSI-PMA structure.

\subsection{Tunability of the magnon-magnon coupling strength}

\begin{figure*}
    \centering
    \includegraphics[width=1\linewidth]{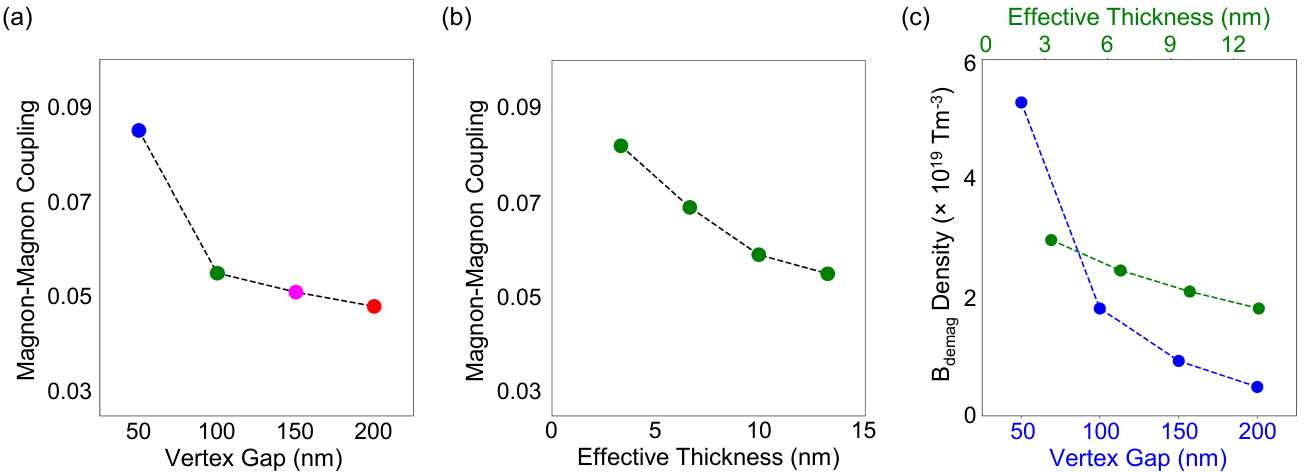}
    \caption{\textbf{Tunable magnon-magnon coupling strength in Type 2}:  The magnon-magnon coupling strength as a function of the vertex gap (a) and the film thickness (b). (c) A combined plot of the magnetostatic field per unit volume as a function of the vertex gap (bottom scale) and the film thickness (top scale).}
    \label{fig4}
\end{figure*}

As indicated in the previous section, the coupling between SSI nanoelements enlarges the frequency gap. Thus, the change of the separation between nanoelements shall also significantly influence hybridization between $\textbf{BM}_{\mathrm{SSI}}$ and $\textbf{BM}_{\mathrm{PMA}}$ modes.  Figure \ref{fig4} (a-c) proves this. Supplementary Information, Fig.~S4 presents the magnetic field-dependent frequency spectrum for SSI-PMA structures with vertex gaps of 50 nm, 150 nm, and 200 nm, while maintaining the same SSI type (Type 2) throughout the analysis. Hybridization between the 2$^\text{nd}$-order $\textbf{BM}_{\mathrm{SSI}}$ and fundamental $\textbf{BM}_{\mathrm{PMA}}$ is observed across all structures in a similar frequency range, but at different magnetic field values. The anticrossing frequency gap varies with the vertex gap: $\Delta f$ = 0.84 GHz for the 200 nm gap, $\Delta f$ = 0.91 GHz for the 150 nm gap, and $\Delta f$ = 1.63 GHz for the 50 nm gap. These gap values are comparable to existing studies on standard ASIs, i.e., without ferromagnetic matrix, where gaps of around 0.3 GHz~\cite{dion2022observation, shen2022dipolar} and 6.5 GHz~\cite{dion2024ultrastrong} have been reported. At smaller vertex gaps, the hybridization between the PMA-SSI boundary region and the bulk of the PMA matrix results in collective oscillations across both areas (Fig.~S4(d-f)). However, as the vertex gap increases, these oscillations become predominantly confined to the nanoelements and the adjacent PMA region, with reduced influence of the bulk part of the PMA matrix, and the dipolar coupling between the nanoelements.

A quantitative measure of the magnon-magnon coupling strength is given by the normalized coupling rate, defined as the ratio of the frequency gap to the higher mode frequency, $\Delta f$/$v$~\cite{dion2024ultrastrong}. Interestingly, decreasing the vertex gap from 200 nm to 50 nm increases the magnon-magnon coupling strength from 0.044 to 0.085 (see, Fig.~\ref{fig4}(a)), at a similar frequency range but at shifted magnetic field values at which anticrossing occurs: 0.40 T for a 50 nm vertex gap, 0.46 T for a 150 nm vertex gap, and 0.49 T for a 200 nm vertex gap (Fig.~S4(a-c)). This coupling enhancement is primarily attributed to the closer proximity and higher density of nanoelements in the PMA matrix. Thus, the integration of the ferromagnet\cite{Qin2018} and the arrangement of a dense array of nanoelements significantly enhances the traditionally weak dipolar interactions, resulting in significant dynamic effects and strong coupling. It is noteworthy that these dynamic effects are correlated with the change of the static magnetization at the edges of the nanoelements. They form an S-shaped configuration at a small vertex gap due to the strong dipolar attraction between them and as the vertex gap increases, this S-shaped curvature of the magnetization diminishes [see the inset in Fig.~S4 (a-c) in SI]. 

To further understand the effect of the dipolar field on the magnon-magnon coupling in the SSI-PMA system, we present the plot of the coupling strength as a function of the film thickness in Fig.~\ref{fig4} (b) with the vertex gap of 100 nm. The applied field versus frequency spectrum and the hybridized modes are shown in the SI, Sec.~V. Interestingly, magnon-magnon coupling strength gradually increases from 0.055 to 0.082 with decreasing the thickness values from 13.2 nm to 3.3 nm ($\Delta f$ = 1.04, 1.10, 1.24 and 1.36 GHz for the 13.2,  9.9,  6.6, and  3.3 nm thickness values, respectively). This is an unexpected result since increasing the thickness effectively increases the magnetostatic field in the system, and one might also expect an increase in the dipolar interactions. To isolate the effect of dipolar interactions within and between nanoelements, we normalized the magnetostatic field by the volume of the system,  thereby removing the dependence on both thickness and vertex gap.  In Fig. \ref{fig4}(c), we plot the magnetostatic field density as a function of vertex gap at a fixed thickness of 13.2 nm, and as a function of the film thickness at a vertex gap of 100 nm. These results indicate that decreasing the vertex gap increases the magnetostatic field density due to enhanced dipolar coupling between the nanoelements, which, in turn, leads to a more pronounced dynamic magnon-magnon hybridization. Conversely, reducing only the effective thickness at a constant vertex gap results in an increased internal magnetic field within the nanoelements~\cite{li2017thickness}, which is reflected in the increased magnetostatic field density and stronger coupling between the magnons.

\subsection{Magnon-magnon hybridization in a ground and monopole vertex configurations \label{Sec:Monopole}}

\begin{figure*}
    \centering
    \includegraphics[width=1\linewidth]{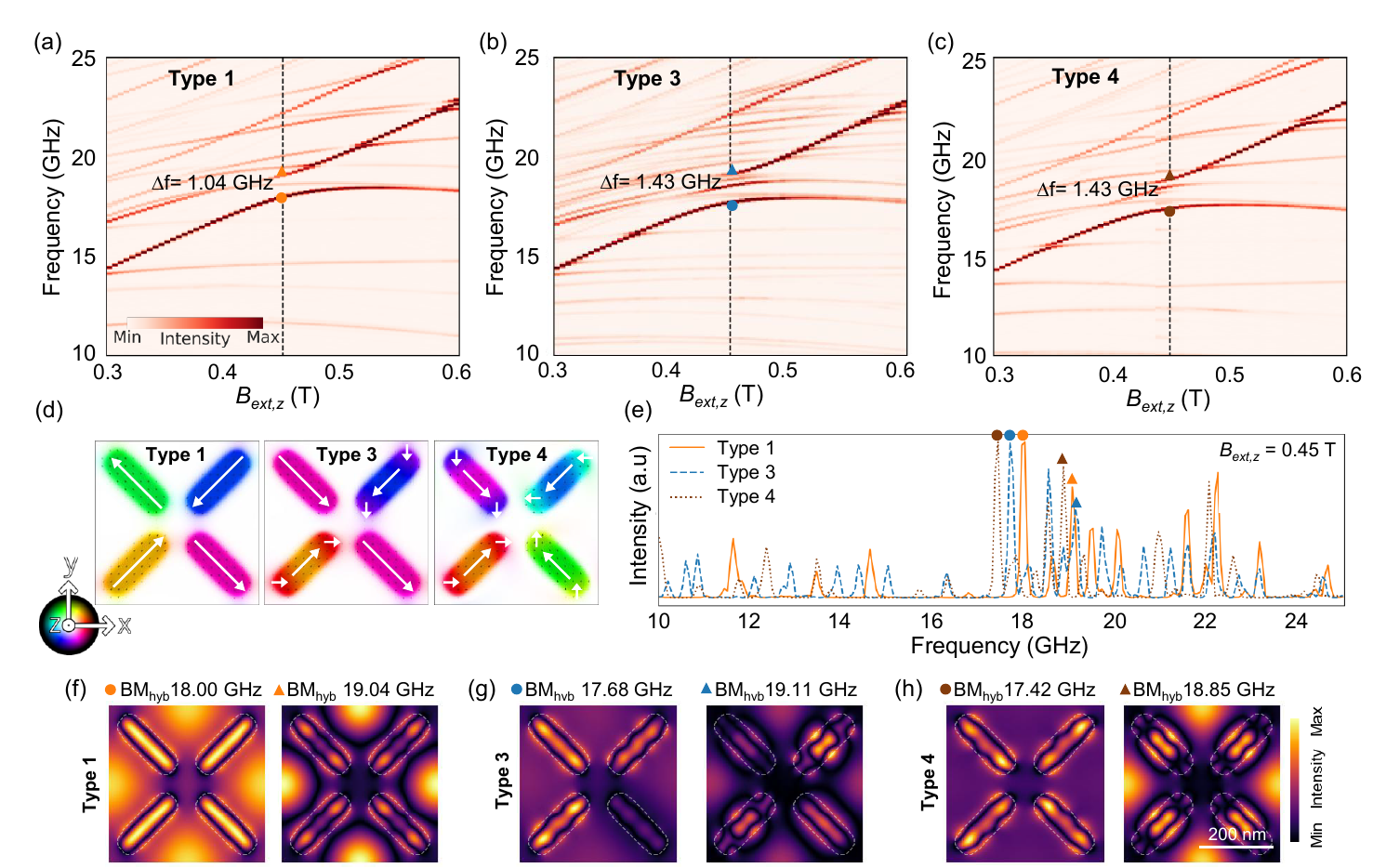}
    \caption{\textbf{SSI Type dependent mode hybridization and anticrossings}:  The zoomed-in section of bias magnetic field-dependent frequency spectra for Type 1 (a), Type 3 (b), and Type 4 (c) vertex configuration, and (d) the corresponding magnetization configuration at 0.05 T. Vertical dashed lines indicate the anticrossing frequency gaps at $B_{ext,z} = 0.45$ T, with gap widths of 1.04 GHz for Type 1 and 1.43 GHz for Type 3 and Type 4. (e) FFT spectra for $B_{ext,z}$ = 0.45T for Type 1, Type 3 and Type 4 configurations. (f), (g) and (h) The snapshots of the dynamic magnetization profiles of the coupled magnons at the anticrossing [marked in (a)-(c)] for Type 1, Type 3 and Type 4, respectively. For improved visibility in $\textbf{BM}_{\mathrm{hyb}}$, the contours of the in-plane nanoelements are highlighted with dotted lines. The shown modes are marked with full dots in the spectra (a)-(c) and (e).}
    \label{fig5}
\end{figure*}

As introduced in Fig.~\ref{fig1}, in the SSI-PMA, as well as in classical square lattice ASIs ~\cite{gliga2020dynamics, farhan2013direct, zhang2019understanding}, the magnetization configuration at the vertex can also be in Type 1, i.e., the lowest energy state in ASIs, or at high energy states Type 3 or Type 4. These possibilities are promising for the reconfigurability of the system for magnonic applications~\cite{gliga2013spectral, gartside2021reconfigurable, mondal2024brillouin}. In this section, we show how changing the vertex type in the unit cell of SSI-PMA affects the magnon-magnon hybridization found in the system with Type 2 vertices. 

The frequency spectrum versus external field and selected SW mode profiles for SSI-PMA with Type 1, Type 3, and Type 4 magnetization configurations in the unit cell are shown in Fig.~\ref{fig5}.  In all cases, the vertex gap is 100 nm, and the corresponding spectrum for Type 2 is shown in Fig.~\ref{fig3}. The selected magnetization state in nanoelements is artificially introduced into the simulations, after which the system is relaxed, according to the procedure detailed in SI, Sec.~I. The spectra for Types 1 and 2 show very similar trends in mode hybridization [Fig.~\ref{fig5}(a)], with only a slight shift of the modes to higher frequencies (about~0.13 GHz for the hybridized modes [Fig.~\ref{fig5}(f) and [Fig.~\ref{fig3}(b)]). The nature of the EM and $\textbf{BM}_{\mathrm{SSI}}$ in the Type 1 configuration [Fig.~S7(a) in SI] is the same as observed in the Type 2 configuration (Fig.~\ref{fig2}). In contrast, Types 3 (monopole) and 4 (high-energy monopole state) [Fig.~\ref{fig5}(b) and (c), respectively] exhibit different spectra and associated SW amplitude distributions, which will be discussed in detail in the following part of the section.

Unlike in Type 1 or Type 2, in Type 3 the static magnetization configurations in nanoelements are significantly modified, forming the S-state in two of the nanoelements, as depicted in Fig.~\ref{fig5}(d), pointing at the strong dipolar coupling between the nanoelements. This leads to frequency splitting of the oscillations at the edges and bulk of the individual nanoelements [see the spectra in Fig. S7(b)]. In the Type 4 configuration, the magnetization of all nanoelements follows an S-state with pronounced curvature at the edges, forming a vortex-like state at the vertex [in Fig.~\ref{fig5}(d) it is oriented counterclockwise]. This leads to a reduction of the demagnetization field compared to other vertex configurations and splitting of the EM and $\textbf{BM}_{\mathrm{SSI}}$ modes [Fig. S7(c)]. The selected profiles of low-frequency modes, EMs, and $\textbf{BMs}_{\mathrm{SSI}}$ at 0.25 T field for Type 1 Type 3, and Type 4 are shown in Fig. S7(d)-(f). In contrast to the system obeying the ice rule, where excitations are typically collective from all nanoelements, monopole-type systems show rather isolated excitations of the nanoelements, as in Type 3. In the Type 4 structure, however, the EMs cover a pair of nanoelements, and the BMs are excited in all four, but in the orthogonal pairs are of different types.

The strong hybridization observed in Types 1 and 2 also exists in the high-energy monopole states. In Type 3 we observe a strong hybridization between second-order $\textbf{BM}_{\mathrm{SSI}}$ and fundamental mode of PMA but in oppositely coupled nanoelements, as shown in Fig.~\ref{fig5}(g), maintaining a significant anticrossing frequency gap $\Delta f=1.43$~GHz at 0.45 T. Remarkably, for Type 4, a collective mode hybridization is observed in Fig.~\ref{fig5}(h) with the anticrossing frequency gap width similar to Type 3 (see also the phase distribution in SI Fig. S9).

The estimated magnon-magnon coupling strength for Type 1 is 0.055, while for the monopole states, it is 0.075. Previous studies have already shown that SW mode hybridization and an anticrossing gap in ASIs composed of the bilayered nanodots are enhanced, e.g., gap width increases from 0.22 GHz to 0.30 GHz in transitioning from ground to monopole state~\cite{gartside2021reconfigurable}. Our results indicate that when square ice is embedded in a PMA matrix and both dipolar and exchange interactions play an important role, the monopole state also further enhances the magnon-magnon coupling strength. 

It is worth noting that several magnonic modes are present within the anti-crossing frequency gap, which is particularly evident in Type 3. Such multimode coupling is quite common, especially at higher frequencies, in magnetic systems with complex structures and magnetization textures~\cite{ghirri2023ultrastrong}. The strongest modes are assumed to be dominant. We follow the same approach and consider the high amplitude mode to have strong coupling to the fundamental mode of the PMA, while other modes have weaker coupling. Furthermore, these strongest coupled modes should be visible in standard broadband ferromagnetic resonance measurements for both in-plane and out-of-plane microwave magnetic field excitations associated with the microstrip (or coplanar microwave) or ring type of microwave antenna, respectively, as shown in SI Fig. S8.

\section{Conclusion}

We investigate the SW dynamics and mode hybridization in a square ASI system (SSI) immersed in the ferromagnetic matrix with PMA (SSI-PMA), revealing complex spectral compositions and unique dynamics that are absent in standalone SSI systems. We show that embedding the SSI in the PMA matrix introduces additional SW modes and interactions into the system, leading to hybridization between the bulk modes of the SSI nanoelements and the fundamental bulk modes of the PMA, resulting in substantial anticrossing frequency gaps. In addition to the previously reported dipolar coupling in standard ASIs, we demonstrate that exchange interactions are crucial for facilitating mode hybridization in SSI-PMA system. For the ground state vertex configurations of the magnetization in nanoelements (Types 1 and 2), we observed an anticrossing frequency gap of 1.04 GHz, which points to the magnon-magnon coupling strength of 0.055. This can be adjusted by changing the vertex gap between the nanoelements and also the effective thickness of the structure. Importantly, we show that the strength of this magnon-magnon coupling can also be modulated by reconfiguring the magnetization orientation in SSI nanoelements.  In particular, we show that high monopole vertex states (Types 3 and 4) of SSIs exhibit different spectra from low energy states with an enhanced anticrossing frequency gap of 1.43 GHz, giving a coupling strength of 0.075, i.e., 36\% enhancement. 

Although the proposed structure is based on Co/Pd multilayers with in-plane magnetized nanoelements resulting from the destruction of anisotropy by ion bombardment, the simulations show that, assuming rational, relatively high damping and using standard broadband FMR measurements, the described coupling should be measurable. Further improvement can be achieved by replacing metallic multilayers of PMA with low-damping materials, e.g. dopped YIG~\cite{das2024tuning}.

Our results demonstrate that the SSI-PMA system is a new kind of ASI with enhanced dynamical magnetic interactions, promising for magnonic applications and further exploration ASI for SW propagation. Given the importance of recent advances in vertex-state dependent neuromorphic computing capabilities within bi-layered ASI systems~\cite{dion2024ultrastrong, manneschi2024optimising}, our findings offer a promising new approach for realising similar systems without patterning, but with enhanced functionality.

\begin{acknowledgments}
The study has received financial support from the National Science Centre of Poland, Grant Nos.~UMO-2020/37/B/ST3/03936 and 2023/49/N/ST3/03538. The simulations were partially performed at the Poznan Supercomputing and Networking Center (Grant No.~PL0095-01). 
\end{acknowledgments}
\section{References}
\bibliographystyle{ieeetr}
\bibliography{bibliography}


\end{document}


\title{Supplementary material: Enhancement of dynamical coupling in artificial spin-ice systems by incorporating perpendicularly magnetized ferromagnetic matrix}

\author{Syamlal Sankaran Kunnath}
    
\author{Mateusz Zelent}

\author{Mathieu Moalic}

\author{Maciej Krawczyk}
  
\affiliation {Institute of Spintronics and Quantum Information, Faculty of Physics and Astronomy, Adam Mickiewicz University, Poznan, 61-614 Poznan, Poland}

\date{\today}
\maketitle

\section{Micromagnetic simulations}

The micromagnetic simulations are performed using own version of $Mumax^{3}$, called $Amumax$, which solves numerically the Landau--Lifshitz--Gilbert equation:
\begin{equation}
 \frac{\text{d}\mathbf{m}}{\mathrm{d}t}= 
 \frac{\gamma \mu_0}{1+\alpha^{2}} \left(\mathbf{m} \times \mathbf{H}_{\mathrm{eff}} + 
 \alpha  \mathbf{m} \times 
 (\mathbf{m} \times \mathbf{H}_{\mathrm{eff}}) \right),
\end{equation}
where $\textbf{m} = \textbf{M} / M_{\mathrm{S}}$ is the normalized magnetization, $\textbf{\text{H}}_{\mathrm{eff}}$ is the effective magnetic field acting on the magnetization, $\gamma=187$ rad/(s$\cdot$T) is the gyromagnetic ratio, $\mu_0$ is the vacuum permeability and $\alpha$ is gilbert damping.
The following components are considered in $\textbf{H}_{\mathrm{eff}}$: demagnetizing field $\textbf{\text{H}}_{\mathrm{d}}$, exchange field $\textbf{\text{H}}_{\mathrm{exch}}$, uniaxial magnetic anisotropy field $\textbf{\text{H}}_{\mathrm{anis}}$, and external magnetic field $\textbf{\text{H}}_{\mathrm{ext}}$, and thermal effects have been neglected:
\begin{equation}
  \textbf{H}_{\mathrm{eff}} =
  \textbf{H}_{\mathrm{d}} + \textbf{H}_{\mathrm{exch}} + \textbf{H}_{\mathrm{ext}} + \textbf{H}_{\mathrm{anis}} +\textbf{h}_{\mathrm{mf}}.
\end{equation}
The last term, $\textbf{h}_{\mathrm{mf}}$ is a microwave magnetic field used for spin wave (SW) excitation. The exchange and anisotropy fields are defined as
\begin{equation}
  \textbf{H}_{\mathrm{exch}} = \frac{2A_{\mathrm{ex}}}{\mu_0 M_{\mathrm{S}}} \Delta \textbf{m},\;
  \textbf{H}_{\mathrm{anis}} =
\frac{2K_{\mathrm{u,bulk}}}{\mu_0 M_{\mathrm{S}}} m_z \hat{\textbf{z}},
\label{Eq:Fields}
\end{equation}
where $A_{\mathrm{ex}}$ is the exchange constant, $M_{\mathrm{S}}$ is the saturation magnetization, and $K_{\mathrm{u,bulk}}$ is bulk magnetic anisotropy constant.

The unit cell of the system is discretized into a $128 \times 128$ grid of cuboids in the $x$ and $y$ directions, with each unit cell measuring approximately $3.31 \times 3.31 \times$ [Co/Pd thickness]~nm$^3$. The cell size adapts according to the lattice constant change later but always remains smaller than the exchange length, 5.62 nm to ensure accuracy. Periodic boundary conditions with 16 repetitions are applied in both the $x$ and $y$ directions to form the 2D lattice instead of a single unit cell.

The stabilization of the magnetization configuration is obtained in the following steps. For the SSI-PMA structure, initially, the magnetization of each cell in the PMA bulk part is aligned along the $z$-axis, while the SSI nanoelements are kept along the x-axis. After the relaxation, the SSI stabilizes into Type 2 vertex state. Two approaches were employed: first, for the hysteresis loop simulation, the external magnetic field is swept along the out-of-plane direction. For each applied field value, the system is relaxed to its corresponding equilibrium state by integrating the LLG equation. The distinct magnetization configurations, specifically Types 1, 3, and 4, were achieved by initializing the magnetization in a predetermined arrangement corresponding to the desired type, followed by a relaxation process. For the dynamic simulations, the system is first relaxed to a ground state under the given applied field values. Subsequently, excite the SWs with a global, spatially homogeneous microwave magnetic field along the $x$-axis with a $sinc$ temporal profile, a cut-off frequency of 25 GHz, and a peak amplitude of $5 \times 10^{-4}$~T. The excitation field is applied for 1.6 ns, and we sample the magnetization dynamics at intervals of 15.38~ps over a period of 15.38~ns.

To obtain the SW spectrum, we took the space- and time-resolved magnetization, applied a Hanning window along the time axis, and computed the real discrete Fourier transform using the Fast Fourier Transform (FFT) algorithm along the time axis for each cuboid composing the system. After this process, the cell with the maximum amplitude was located for each discrete frequency.  This step was repeated iteratively for each frequency to progressively construct the spectrum. 
By selecting the highest amplitude rather than the average, the strongly-localized mode is emphasized over the modes that are spread over a wide area, such as the bulk modes. This procedure is applied to each simulation, where the value of the external magnetic field varies. The SW spectral response as a function of the external magnetic field directed along the $z$-axis is calculated for values from 0.05 to 1 T in increments of 25 mT. For the selected magnetic field range, the simulations are performed with 5 mT increments. To prevent numerical artifacts due to the supersymmetry of the system, we angle the external field by 0.0001 degrees from the $z$-axis. To visualize the modes, we independently processed each cuboid within the system by calculating the FFT of the in-plane magnetization over time. For a selected frequency, the modulus of the resulting complex number is mapped to a saturation value ranging from 0 to 1, while the argument is mapped to a hue, with red representing an argument of 0. 

\section{Spectrum of PMA matrix with the holes}
\begin{figure*}
    \centering
    \includegraphics[width=0.6\linewidth]{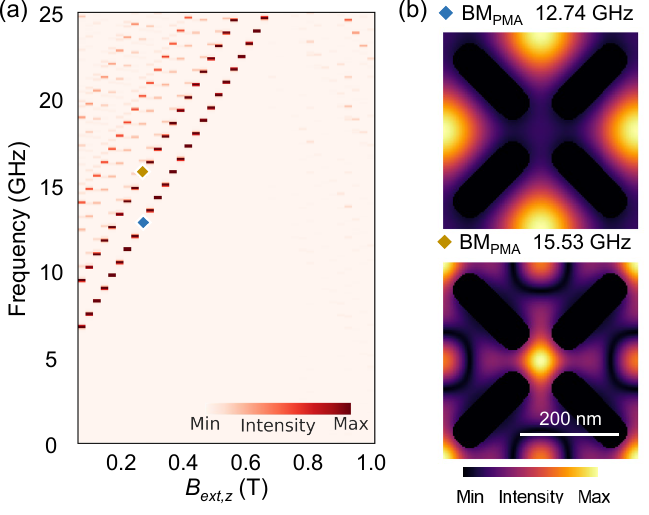}
    \caption{(a) Evolution of the SW resonance spectra for the PMA matrix with the holes instead of nanoelements in dependence on the static external magnetic field directed out-of-plane, (b) snapshots of most intensive SW modes at 0.25 T field. The modes are highlighted in the spectra. The frequencies of the modes are marked on the (a) plot.}
    \label{S1}
\end{figure*}

Figure ~\ref{S1} presents the external magnetic field dependence of the SW spectra [Fig.~\ref{S1}(a)] and highlights the most intensive modes [Fig.~\ref{S1}(b)] for the structure consisting solely of the PMA matrix (with the holes instead of nanoelements).
 Within this system, we observe the fundamental bulk mode ($\textbf{BM}_{\mathrm{PMA}}$) as well as higher-order $\textbf{BM}_{\mathrm{PMA}}$, where the SWs are excited at the bulk region of the PMA matrix ($\textbf{BM}_{\mathrm{PMA}}$) and both at the vertex gap and bulk part (higher order $\textbf{BM}_{\mathrm{PMA}}$). For instance, at the external magnetic field of 0.25 T, the $\textbf{BM}_{\mathrm{PMA}}$ appears at 12.74 GHz, and the second-order $\textbf{BM}_{\mathrm{PMA}}$ is observed at 15.53 GHz [see Fig.~\ref{S1}(b)]. Increasing the magnetic field results in a monotonic increase in the frequency of these modes, in contrast to the SSI modes (Fig.~2 (a), (b)).

\section{Role of the exchange interactions in magnon-magnon coupling}
\begin{figure*}
    \centering
    \includegraphics[width=1\linewidth]{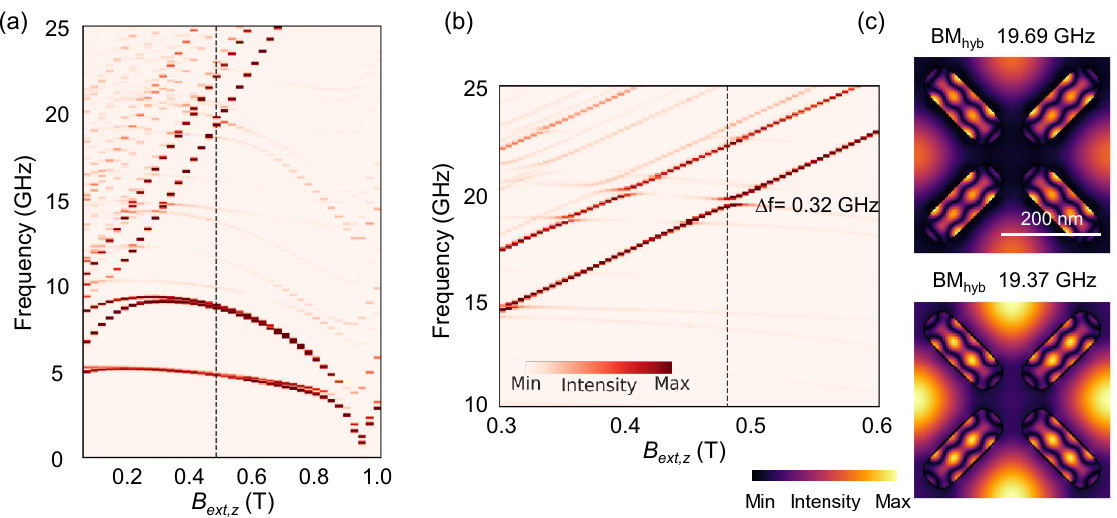}
    \caption{The evolution of SW resonance spectra in dependence on the applied magnetic field (a), the zoomed-in section of the spectra at higher frequency range (b), and the hybridized modes (c) for an SSI-PMA system that includes a thin non-magnetic spacer layer surrounding all the nanoelements to break the exchange interaction between subsystems.}
    \label{S2}
\end{figure*}

To understand the origin of the observed hybridization between the $\textbf{BM}_{\mathrm{PMA}}$ and second-order $\textbf{BM}_{\mathrm{SSI}}$ in SSI-PMA systems (referred to as $\textbf{BM}_{\mathrm{hyb}}$ for the hybridized modes), we performed simulations incorporating a non-magnetic spacer layer with a dimension of approximately 6.62 nm (2 cell sizes) between the PMA matrix and the nanoelements. This breaks the exchange coupling between ferromagnetic nanoelement and the PMA matrix. The frequency versus field spectra (Fig. \ref{S2}(a, b)) revealed a significant reduction in the anticrossing frequency gap from 1.04 GHz (SSI-PMA without the spacer layer, Fig.~3 of the main text) to 0.32 GHz.  This decrease in the frequency gap indicates that the hybridization between $\textbf{BM}_{\mathrm{PMA}}$ and the second-order $\textbf{BM}_{\mathrm{SSI}}$ modes is weak when a non-magnetic spacer is present (Fig. \ref{S2}(c)). As the spacer layer disrupts the direct exchange interactions between the PMA matrix and the nanoelements and weakens the magnon-magnon coupling, it highlights the critical role of exchange interactions in facilitating strong mode hybridisation and dynamic coupling in our magnonic systems.  

\section{Role of the inter-nanoelement interactions on magnon-magnon coupling}
\begin{figure*}
    \centering
    \includegraphics[width=1\linewidth]{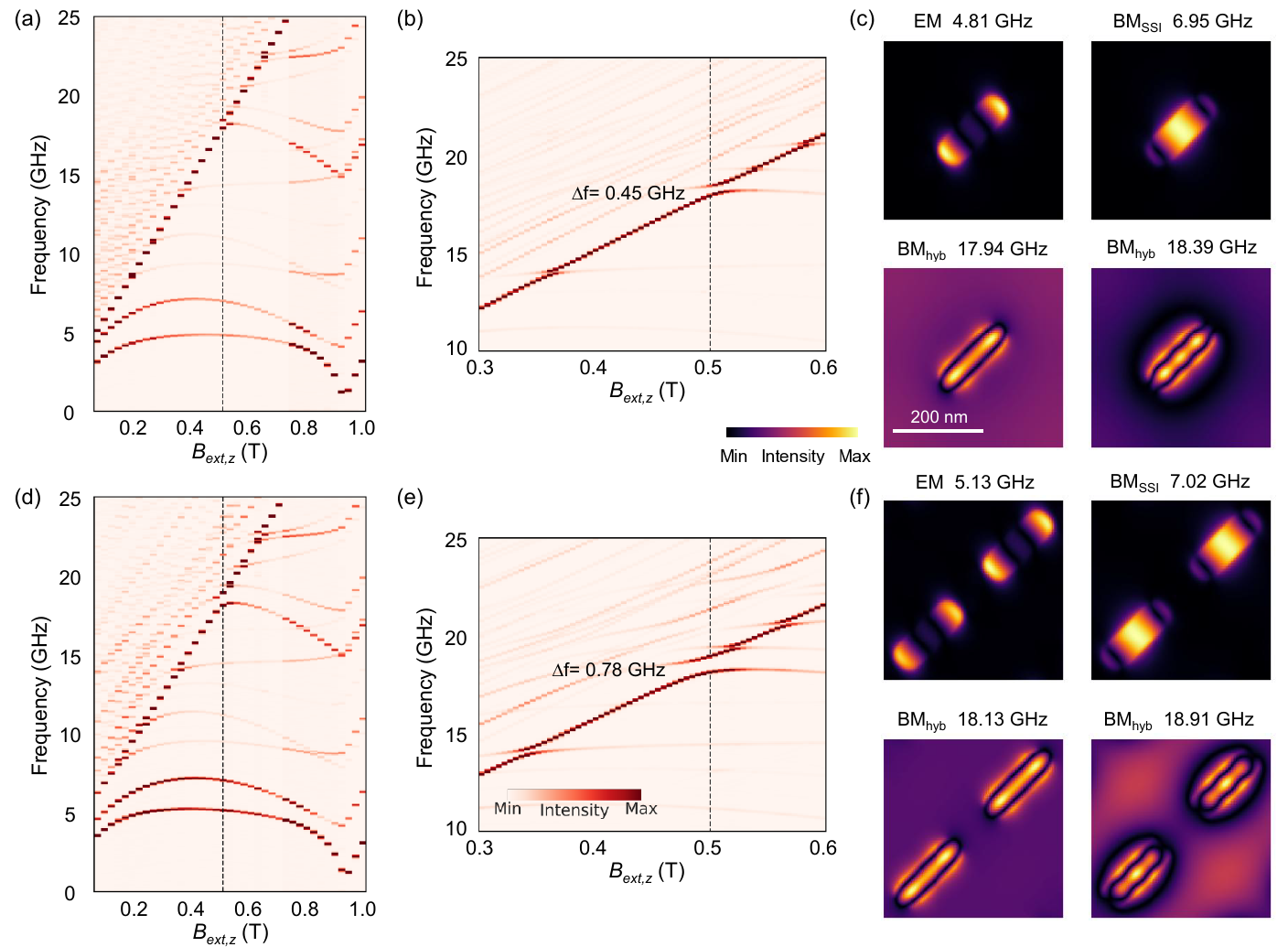}
    \caption{The evolution of the SW spectra for a single nanoelement (a) and two nanoelements (d) in the unit cell embedded in a PMA matrix. The zoomed section of these spectra is shown in (b) and (e), respectively. In the single-element-PMA system, an anticrossing frequency gap of 0.45 GHz is observed, whereas the two-element-PMA system exhibits a gap of 0.78 GHz both at 0.5 T field. (c) and (f) The mode profiles for the EM, BM, and hybridized modes at an applied field of 0.5 T for the single-element-PMA and two-element-PMA systems, respectively.}
    \label{S3}
\end{figure*}

To elucidate the role of the interactions between the ferromagnetic in-plane magnetized nanoelements in the studied systems, we conducted simulations on two different configurations where in each unit cell there is: (i) a single nanoelement in a PMA matrix, where the interelement interactions are absent, and (ii) there are two-nanoelements with a 100 nm gap in a PMA matrix, where interelement dipolar interactions are expected. The external field-dependent frequency spectra and mode profiles for the single nanoelement-PMA system are shown in Fig. \ref{S3}(a,b), while the results for the two-nanoelement system are presented in Fig. \ref{S3} (d,e). Interestingly, both configurations exhibit hybridizatiion between $\textbf{BM}_{\mathrm{PMA}}$ and the second-order $\textbf{BM}_{\mathrm{SSI}}$ leading to an anticrossing frequency gap in their spectra. However, the magnitude of this gap varies significantly between the two systems. For the single nanoelement-PMA system, $\Delta f = 0.45$ GHz, and for the two-nanoelement-PMA system there is an enhanced anticrossing frequency gap, i.e.,  $\Delta f = 0.78$ GHz. The mode profile shown in Fig. \ref{S3} (c,f) provides further insight into the differences between the systems. In the single nanoelement-PMA system, the observed mode hybridization is due to the direct exchange and dipole interaction between the PMA matrix and the nanoelement, as its value is slightly higher than the gap in Type 2 with the braking exchange interactions (0.32 GHz, see Fig.~ \ref{S2}). In the two-nanoelement-PMA system, the second-order $\textbf{BM}_{\mathrm{SSI}}$ of dipolar coupled nanoelements collectively hybridized with $\textbf{BM}_{\mathrm{PMA}}$ resulting in an enhanced $\Delta f$ by 0.33~GHz, and the magnon-magnon coupling strength from 0.024 to about 0.041. These findings emphasize the critical role of interelement interactions, most probably dipolar, in enhancing mode hybridization and magnon-magnon coupling strength. This interelement coupling is responsible for the significant enhancement of the magnon-magnon coupling strength to 0.055 in a full Type 2 SSI-PMA system (Fig. 3).  

In conclusion, while the exchange interaction between the PMA matrix and individual nanoelements establishes the basic mode hybridization, the presence of additional nanoelements and the resulting interelement dipolar interactions significantly enhance the coupling effects.

\section{SW spectra in Type 2 SSI-PMA system in dependence on the vertex gap and effective thickness}
\begin{figure*}
    \centering
    \includegraphics[width=1\linewidth]{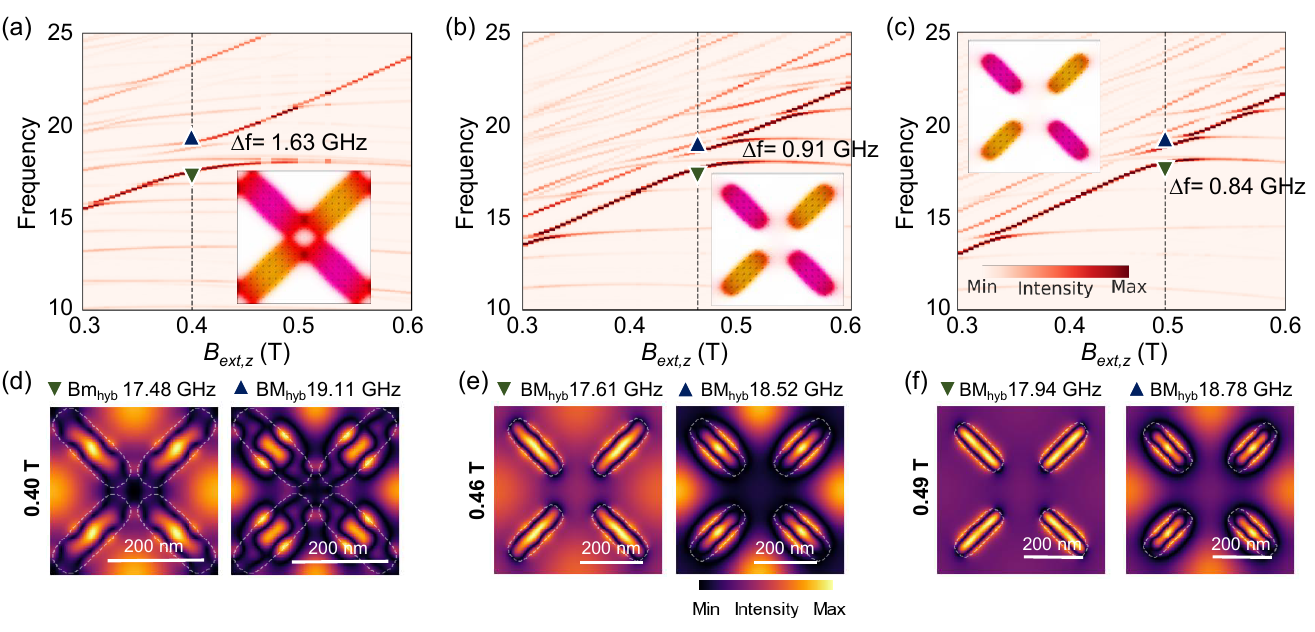}
    \caption{The zoomed section of the frequency versus magnetic field spectra for SSI-PMA in the Type 2 state with the vertex gaps of 50 nm (a), 150 nm (b), and 200 nm (c). The magnetic fields were swept from 0.3 to 0.6 T, with a step size of 5 mT. The insets show the static magnetization configuration at the fields marked by the vertical dashed line, indicating the anticrossing frequency gaps. (d), (e) and (f) The snapshots of the dynamic magnetization profiles of the coupled magnons at the smallest frequency gap field for the vertex gaps of 50 nm, 150 nm, and 200 nm, respectively.}
        \label{S4}
\end{figure*}
\begin{figure*}
    \centering
    \includegraphics[width=1\linewidth]{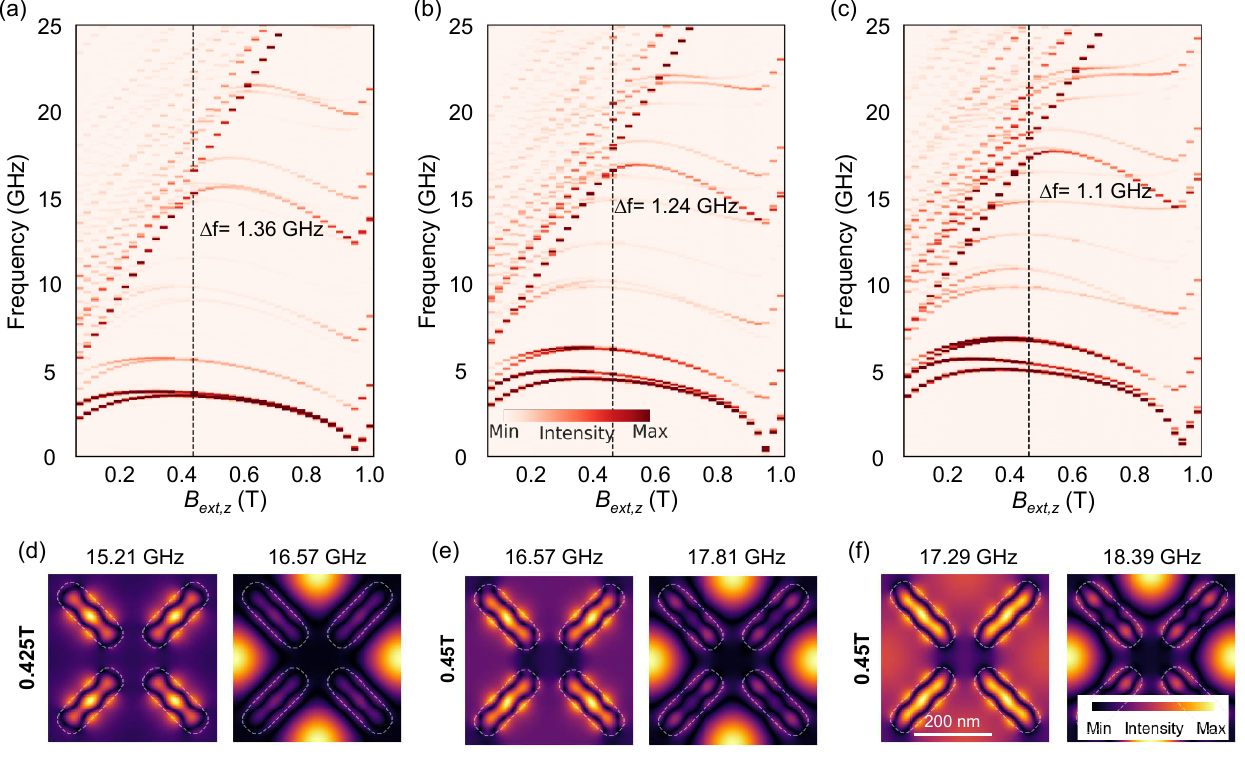}
    \caption{The frequency versus magnetic field spectra for SSI-PMA Type 2 configurations with the effective thickness values of 3.3 nm (a), 6.6 nm (b), and 9.9 nm (c). The observed anticrossing frequency gaps decrease with increasing thickness, observed at 1.36 GHz for 3.3 nm, 1.24 GHz for 6.6 nm, and 1.10 GHz for 9.9 nm. Panels (d-f) show the SW profiles of the modes involved in hybridisation for each thickness.}
    \label{S5}
\end{figure*}

Figure \ref{S4} (a-c) shows the magnetic field-dependent frequency spectrum of SWs in SSI-PMA structures with vertex gaps of 50 nm, 150 nm, and 200 nm, while maintaining the same SSI type (Type 2) throughout the analysis. Hybridization between the 2$^\text{nd}$-order $\textbf{BM}_{\mathrm{SSI}}$ and the fundamental $\textbf{BM}_{\mathrm{PMA}}$ is observed for all structures in a similar frequency range, but at different magnetic field values. The anticrossing frequency gap varies with the vertex gap: $\Delta f$ = 0.84 GHz for the 200 nm gap at 0.50 T, $\Delta f$ = 0.91 GHz for the 150 nm gap at 0.475 T, and $\Delta f$ = 1.63 GHz at 0.40 T for the 50 nm gap. The corresponding magnon-magnon coupling strength as a function of thickness can be found in Fig. 4(a) of the main manuscript. 

Figure \ref{S5} presents the magnetic field-dependent frequency spectra and mode profiles for SSI-PMA systems with three different effective thicknesses of the multilayer, without affecting their magnetization state, i.e., keeping the Type-2 magnetization configuration. As detailed in the Methods section, we used a Co(0.75 nm)/Pd(0.9 nm) multilayered system as the basis for this study. To systematically explore the role of thickness in mode hybridization and magnon coupling, simulations were performed for three different effective thicknesses: 3.3 nm (2 repetitions of Co/Pd) (Fig. \ref{S5} (a, d)), 6.6 nm (Fig. \ref{S5} (b, e)), and 9.9 nm (Fig. \ref{S4} (c, f)), with a constant vertex gap of 100 nm between the SSI nanoelements. The results for an effective thickness of 13.2 nm are presented in Fig. 2 of the main manuscript. Across all the studied thicknesses, hybridization between the second-order $\textbf{BM}_{\mathrm{SSI}}$ and $\textbf{BM}_{\mathrm{PMA}}$ is consistently observed. However, a notable trend emerges in the anticrossing frequency gap ($\Delta f$), which decreases with increasing thickness: $\Delta f$=1.36 GHz for the 3.3 nm thickness, $\Delta f$=1.24 GHz for the 6.6 nm thickness, and $\Delta f$=1.1 GHz for the 9.9 nm thickness. This indicates that the strength of magnon-magnon coupling is inversely related to the thickness of the system. The corresponding magnon-magnon coupling strength as a function of thickness is depicted in Fig. 4(b) of the main manuscript. Additionally, Fig. 4(c) shows the magnetostatic field density as a function of thickness. Increasing the system's thickness effectively enhances the overall magnetostatic field. However, to specifically isolate the effects of dipolar interactions between the nanoelements, we normalized the magnetostatic field by the volume of the system. This normalization eliminates the dependence on thickness, allowing for a more precise analysis of the interactions. Our results indicate that reducing the effective thickness amplifies the magnetostatic field density, due to intensified dipolar coupling between the nanoelements and leads to more pronounced dynamic magnon hybridization. 

\section{SW spectra in SSI-PMA Type-2 system with enhanced damping}
\begin{figure*}
    \centering
    \includegraphics[width=1.0\linewidth]{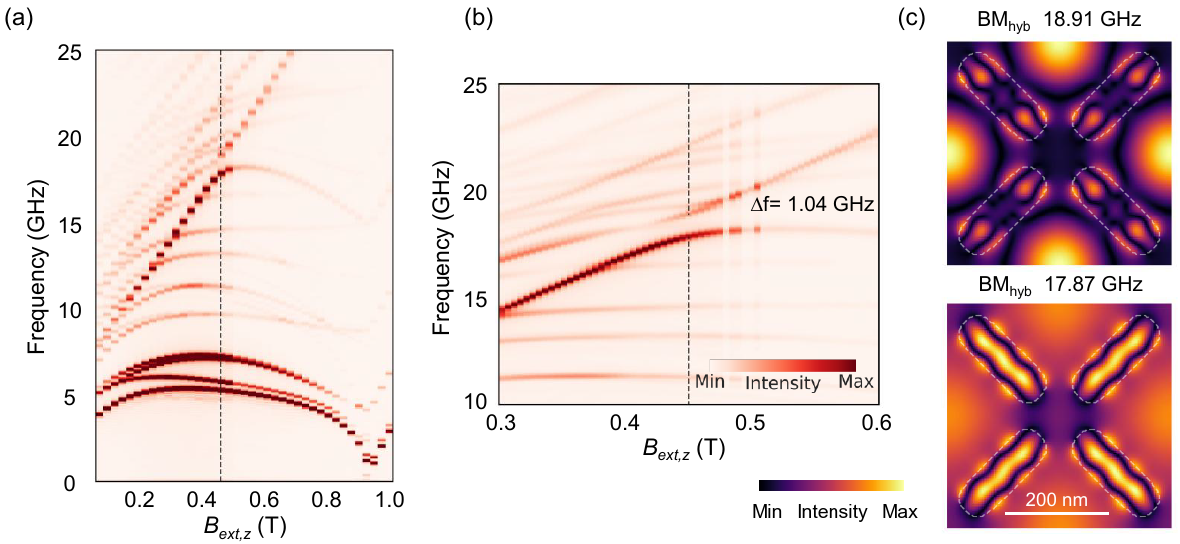}
    \caption{(a) The evolution of the SSI-PMA SW spectra in Type 2 configuration with a damping value of 0.008 as a function of the applied magnetic field. (b) Zoomed-in section of the spectra at a higher-frequency range. The observed anti-crossing frequency gap is 1.04 GHz at 0.45 T. (c) Snapshots of the hybridized modes at 0.45 T.}
    \label{S6}
\end{figure*}

Figure \ref{S6} shows the SW spectra in dependence on the external magnetic field for SSI-PMA structure in Type 2 configuration with a realistic damping value of $\alpha = 0.008$. The results of the same system with a negligible damping value can be found in Fig. 3 of the main manuscript. Notably, the hybridization between $\textbf{BM}_{\mathrm{SSI}}$ and $\textbf{BM}_{\mathrm{PMA}}$ persists under these conditions (Fig. \ref{S6} (b, c)). Compared to Fig. 3, the anticrossing frequency gap and coupling strength remain unchanged.

\section{Influence of the magnetization configuration in the vertex on the magnon-magnon-coupling}
\begin{figure*}
    \centering
    \includegraphics[width=1.0\linewidth]{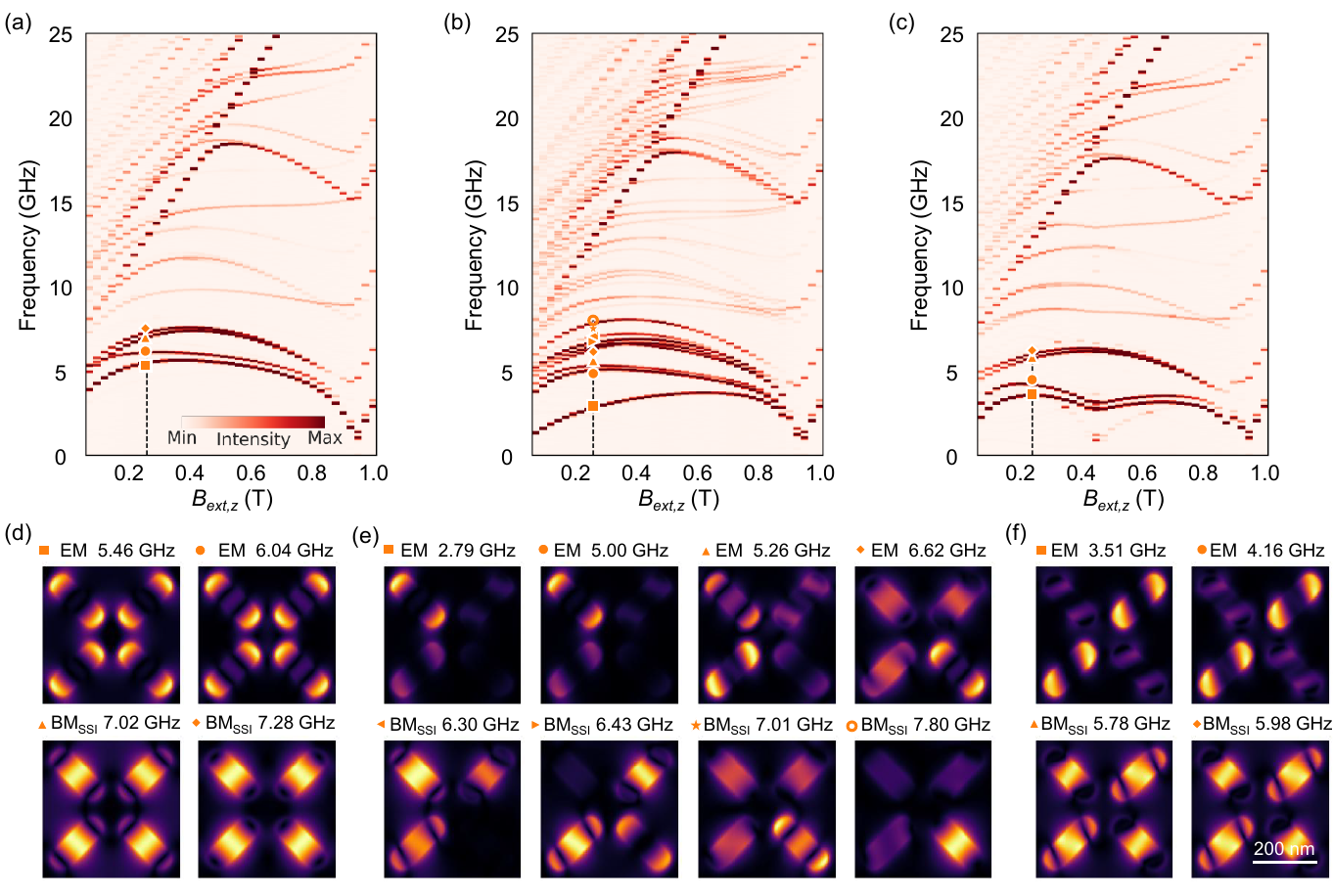}
    \caption{(a-c) Evolution of the SW resonance spectra as a function of the applied magnetic field for Type 1, Type 3 and Type 4 configurations, respectively. The vertical black dashed line indicates the field value 0.25 T in the spectrum. The EMs and $\textbf{BM}_{\mathrm{SSI}}$ corresponding to Type 1 (d), Type 3 (e) and Type 4 (f) configurations at 0.25 T. The modes are clearly marked on the field versus frequency spectra and corresponding images.}
    \label{S7}
\end{figure*}

Figure \ref{S7} (a-c) shows the complete SW resonance spectra as a function of the applied magnetic field for Type 1, Type 3 and Type 4 configurations and the profiles of their EMs and $\textbf{BM}_{\mathrm{SSI}}$ modes at hybridization. The zoomed-in spectra at higher frequency range and mode hybridization observed in these structures are discussed in detail in the main manuscript with the description of Fig.~5. In Fig. \ref{S7} (d-f), the mode profiles are shown at a selected field value of 0.25 T, consistent with the Type 2 profile illustrated in Fig. 2 of the main manuscript. For the Type 1 configurations, the EMs were observed at 5.46 GHz and 6.04 GHz, while the $\textbf{BM}_{\mathrm{SSI}}$ appeared at 7.02 GHz and 7.28 GHz Fig. \ref{S7} (d), where the excitation is collective, similar to the Type 2 observation in Fig. 2 (e). Notably, for Type 3 monopole state, the non-collective excitations result in multiple EMs and $\textbf{BM}_{\mathrm{SSI}}$. Particularly, the EMs are observed at 2.79 GHz, 5.00 GHz, 5.26 GHz and 6.62 GHz and $\textbf{BM}_{\mathrm{SSI}}$ observed at 6.30 GHz, 6.43 GHz, 7.01 GHz and 7.80 GHz as shown in Fig. \ref{S7} (e). For the Type 4 monopole configurations, the orthogonal excitations are observed for EMs at 3.51 GHz and 4.16 GHz, while collective excitation occurs for $\textbf{BM}_{\mathrm{SSI}}$ at 5.78 GHz and 5.98 GHz as shown in Fig.~\ref{S7} (f).

\section{Ferromagnetic resonance spectra of the Type 2 and Type 3 SSI-PMA system}
\begin{figure*}
    \centering
    \includegraphics[width=1.0\linewidth]{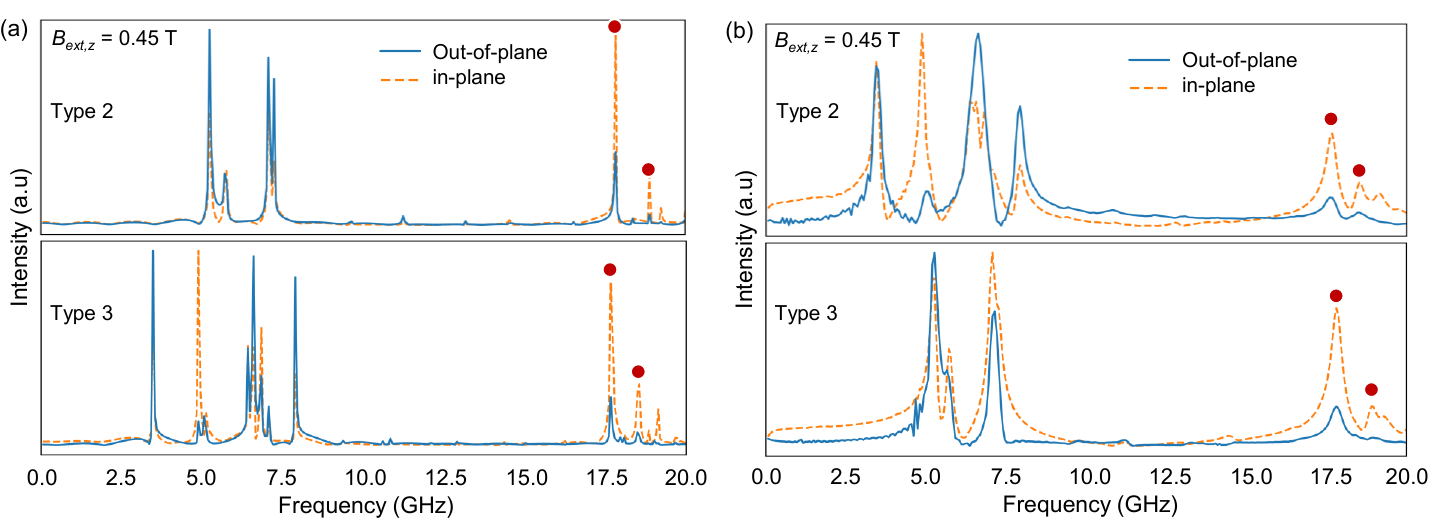}
    \caption{ Average FFT spectra of SSI-PMA at a field value of 0.45~T, illustrating the system response to a homogeneous microwave magnetic field applied along the $x$-axis (orange solid line) and $z$-axis (blue dashed line) for Type 2 and Type 3 configurations with low damping (a) and realistic damping (b) values. The red dots indicate the modes involved in the considered magnon-magnon coupling.}
    \label{S8}
\end{figure*}
\begin{figure*}
    \centering
    \includegraphics[width=0.75\linewidth]{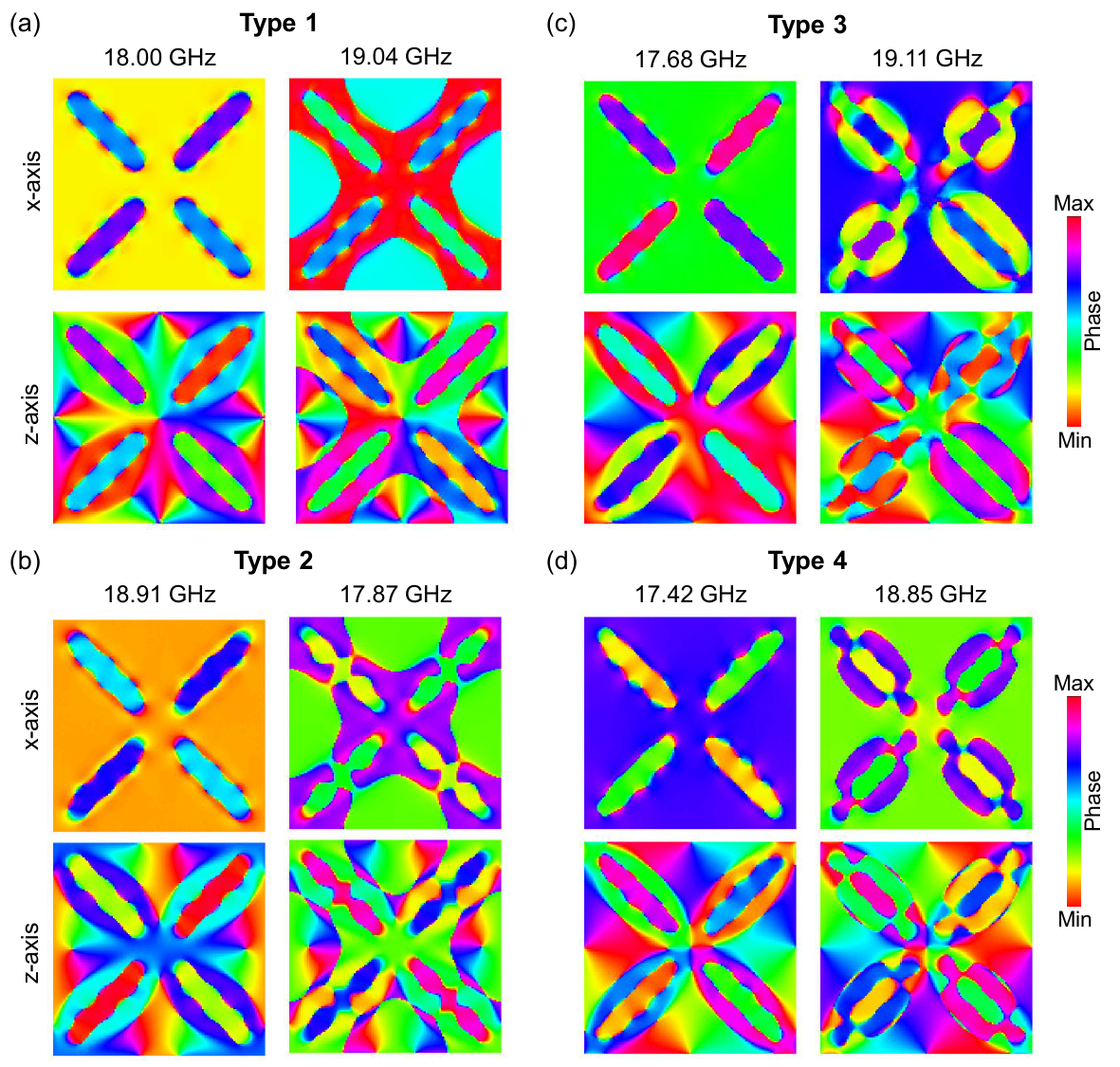}
    \caption{The phase maps of the in-plane ($x$-axis) and out-of-plane dynamic magnetization components of the hybridized modes referred to in the main manuscript: in Fig. 3(b) for type 2 and in Fig. 5(f-h) for Types 1, 3 and 4.}
    \label{S9}
\end{figure*}

We extend our analysis to estimate a possibility for experimental observation of our findings with state-of-the-art broadband ferromagnetic resonance (FMR) measurements, in which a homogeneous microwave magnetic field is applied and the integrated density is proportional to the measured signal. We examine the average FFT spectra for microwave fields oriented along both the $x$- and the $z$-axes, which are related to the two measurement configurations: with the wide microstrip or coplanar waveguide and the ring geometry, respectively.

To calculate the SW spectra, we integrate the oscillating \( i \)-th (the $x$ or $z$) component of the magnetization $( m_i(x, y, z, t))$ over the
 unit cell for each time step:

\begin{equation}
I_i(t) = \int_{\text{unit cell}} m_i(x, y, z, t) \, dx \, dy \, dz,
\end{equation}
Figure \ref{S8}(a) presents these averaged FFT spectra for a biased magnetic field of 0.45 T, contrasting the Type 2 ground state with the Type 3 monopole state configurations in the SSI-PMA system. Our simulations reveal hybridized modes in both configurations, highlighted with red symbols in Fig. \ref{S8}. Additionally, Fig. \ref{S8}(b) presents the corresponding results obtained with a realistic damping value of 0.008. However, for the $x$-axis oriented microwave magnetic field, a significantly stronger absorption is expected due to the more effective excitation of the in-plane magnetized nanoelements in the SSI-PMA structure. The corresponding in-plane magnetization mode profiles are provided in the main manuscript in Fig. 3(b) for Type 2 and Fig. 5(b) for Type 3.

\subsection{Phase maps of the hybridized modes in four different types of SSI-PMA system}

Figure \ref{S9} presents the phase maps of the in-plane ($x$-axis) and out-of-plane dynamic magnetization components of the hybridised modes of SSI-PMA Types 1, 2, 3 and 4. The corresponding amplitude profiles are provided in Fig. 3(b) for Type 2 and in Fig. 5(f), (g), (h) for Types 1, 3, and 4 in the main manuscript.

    
